\documentclass[12pt]{amsart}

\pdfoutput=1

\usepackage[text={420pt,660pt},centering]{geometry}

\usepackage{color}
\usepackage{esint,amssymb}
\usepackage{graphicx,epsfig}
\usepackage{MnSymbol}
\usepackage{mathtools}
\usepackage[colorlinks=true, pdfstartview=FitV, linkcolor=blue, citecolor=blue, urlcolor=blue,pagebackref=false]{hyperref}
\usepackage{microtype}

\usepackage{bm}
\usepackage{scalerel} 
\usepackage[font={footnotesize}]{caption}

\usepackage{mathrsfs}
\usepackage{graphicx}

\usepackage{tikz}
\usepackage{pgf}
\pgfmathsetseed{\number\pdfrandomseed}

\usepgflibrary{arrows} 
\usetikzlibrary{arrows}
\usetikzlibrary{arrows.meta}

\theoremstyle{remark}

\theoremstyle{definition}

\numberwithin{equation}{section}
\numberwithin{proposition}{section}

\newcommand{\R}{\mathbb{R}}
\renewcommand{\le}{\leqslant}
\renewcommand{\ge}{\geqslant}

\newcommand{\E}{\mathbb{E}}
\renewcommand{\P}{\mathbb{P}}

\newcommand{\Zd}{\mathbb{Z}^d}
\newcommand{\Rd}{{\mathbb{R}^d}}

\newcommand{\ep}{\varepsilon}

\renewcommand{\subset}{\subseteq}
\renewcommand{\fint}{\strokedint}
\newcommand{\Ll}{\left}
\newcommand{\Rr}{\right}

\newcommand{\ls}{\lesssim}

\DeclareMathOperator{\var}{var}

\newcommand{\cu}{{\scaleobj{1.2}{\square}}}

\renewcommand{\bar}{\overline}

\newcommand{\mcl}{\mathcal}
\newcommand{\msf}{\mathsf}
\newcommand{\al}{\alpha}

\newcommand{\de}{\delta}

\renewcommand{\a}{\mathbf{a}}
\newcommand{\ahom}{{\overbracket[1pt][-1pt]{\a}}}

\newcommand{\g}{\mathbf{g}}

\renewcommand{\O}{\mathcal{O}}

\renewcommand{\d}{\mathrm{d}}

\begin{document}

\title[An informal introduction to quantitative homogenization]{An informal introduction to quantitative stochastic homogenization}

\begin{abstract}
Divergence-form operators with random coefficients homogenize over large scales. Over the last decade, an intensive research effort focused on turning this asymptotic statement into quantitative estimates. The goal of this note is to review one approach for doing so based on the idea of renormalization. The discussion is highly informal, with pointers to mathematically precise statements.
\end{abstract}

\author[J.-C. Mourrat]{J.-C. Mourrat}
\address[J.-C. Mourrat]{DMA, Ecole normale sup\'erieure,
CNRS, PSL University, Paris, France}
\email{mourrat@dma.ens.fr}


\maketitle

%
%
%
%
%
%

\section{Introduction}

For most physical systems, the model we wish to use for its description depends on the time or space scale of interest. It is then a classical question of mathematical physics to ask whether these different models are consistent with one another as we move from one scale to another.

\smallskip

In this note, we focus on studying the large-scale behavior of divergence-form operators with random coefficients. We expect that, over large scales, an effective differential operator with constant coefficients suffices to capture the coarse properties of the heterogeneous operator. This property can indeed be verified mathematically, in the limit of infinite separation of scales. However, going beyond the verification of this asymptotic consistency and towards explicit rates of convergence, or even towards a precise description of next-order corrections, turns out to be a much more difficult challenge which has been the focus of intensive research over the last ten years. The goal of this note is to provide a gentle introduction to some of the work in this area.

\smallskip

We may choose the language of heat diffusion to describe the equations of interest. For convenience, we study equilibrium problems (that is, without time dependence) posed on a domain of $\Rd$. The equilibrium condition prescribes that the heat flux $\mathbf j$ be of null divergence. Moreover, by Fourier's law, we expect the heat flux to be proportional to the gradient of the temperature $u$, that is, $\mathbf j = - \a \nabla u$. We end up with the classical elliptic equation $-\nabla \cdot (\a \nabla u) = 0$.

\smallskip

The point of focus of this note concerns the situation where the medium in which heat diffuses is inhomogeneous: a typical example to have in mind is that of a composite material. That is, the coefficient field $\a : \Rd \to \R^{d\times d}_\mathrm{sym}$ depends on the space variable (and takes values in the set of symmetric positive definite matrices).  Yet, we do not want to study elliptic equations with arbitrary coefficients. Rather, we think that although the medium is disordered, the statistics of the disorder do not depend on the spatial location. Mathematically, this is encoded in the assumption that the coefficient field $(\a(x), x \in \Rd)$ is \emph{random}, and that its law is invariant under translations. For our purposes, it suffices to assume that this invariance holds under translations along a lattice, which we can fix to be $\Zd$ by a change of variables. 

\smallskip

\begin{figure}
\centering
\begin{tabular}{cc}
\begin{tikzpicture}[scale=0.15]
\draw[lightgray] (0,0) rectangle (30,30);
\foreach \n in {1,2,...,621}
\pgfmathsetmacro{\x}{random(0,29)}
\pgfmathsetmacro{\y}{random(0,29)}
\draw[fill = black] (\x,\y) rectangle (\x+1,\y+1);
\end{tikzpicture}

&

\begin{tikzpicture}[scale=0.45]
\clip (0,0) rectangle (10,10);
\draw[lightgray] (0,0) rectangle (10,10);
\foreach \n in {1,2,...,275}
\pgfmathsetmacro{\x}{(1+rand)*5.5-0.5}
\pgfmathsetmacro{\y}{(1+rand)*5.5-0.5}
\draw[fill = black] (\x,\y) circle (0.2);
\end{tikzpicture}
 
 \vspace{0.5cm}
 \\
\begin{tikzpicture}[scale=0.45]
\clip (0,0) rectangle (10,10);
\draw[lightgray] (0,0) rectangle (10,10);
\foreach \n in {1,2,...,250}
\pgfmathsetmacro{\x}{(1+rand)*5.1-0.05}
\pgfmathsetmacro{\y}{(1+rand)*7-2}
\pgfmathsetmacro{\t}{rand*5}
\pgfmathsetmacro{\H}{(rand+5)*0.1}
\pgfmathsetmacro{\P}{rand*.03}
\pgfmathsetmacro{\Q}{rand*0.015}
\draw[thick,cm={cos(\t) ,-sin(\t) ,sin(\t) ,cos(\t) ,(0 cm,0 cm)}] plot [smooth] coordinates{  (\x,\y)  (\x+\P*\H,\y+\H)  (\x+\P*\H/2 +\Q*\H,\y+2*\H)  (\x,\y+3*\H) };
\end{tikzpicture}

& 
\includegraphics[scale=0.20, trim = 7.5cm 1.5cm 0cm 0cm, clip = true]{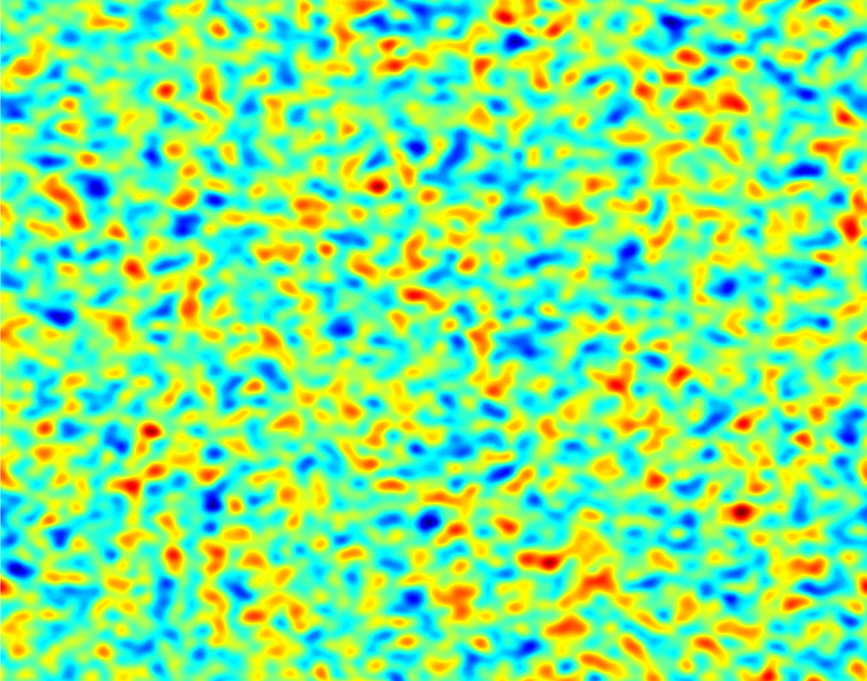}
\end{tabular}
\caption{Examples of coefficient fields, the color scale indicating the value of $\a(x)$. Top left: each unit square is independently colored black or white with probability $1/2$; top right: circles of a fixed radius are drawn around points of a Poisson point process; bottom left: similar to top right but with more complicated shapes; Bottom right: a local function of a white noise field in a color scale.}
	\label{f.coef}
\end{figure}
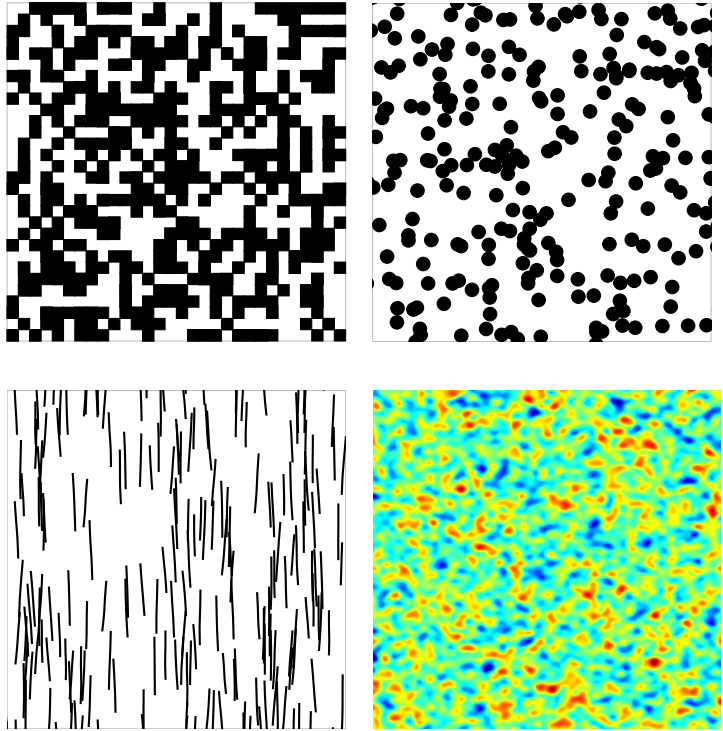

Besides the assumption that the coefficient field is random, we aim to make the simplest possible assumptions to facilitate the presentation of the main ideas. We will thus assume that the coefficient field is uniformly elliptic: there exists a constant $\Lambda \in [1,\infty)$ such that with probability $1$, we have, for every $x \in \Rd$, 
\begin{equation}  
\label{e.unif.ellip}
\forall \xi \in \Rd, \quad \Lambda^{-1}|\xi|^2 \le \xi \cdot \a(x)\xi \le \Lambda|\xi|^2.
\end{equation}
We also assume that the coefficient field has a unit range of dependence: whenever $U,V \subset \Rd$ are separated by a distance at least $1$, the families of random variables $(\a(x), x \in U)$ and $(\a(x), x \in V)$ are independent. Examples of coefficient fields satisfying these assumptions are depicted on Figure~\ref{f.coef}.

In order to instantiate the effect of largely separated scales, we can introduce a small parameter $\ep > 0$ and focus on understanding the solution $u_\ep$ of the problem
\begin{equation}  
\label{e.pde.eps}
\Ll\{
\begin{array}{ll}  
-\nabla \cdot \a \Ll( \tfrac{\cdot}{\ep} \Rr) \nabla u_\ep = 0  & \quad \text{in } U ,\\
u_\ep = f & \quad \text{on } \partial U,
\end{array}
\Rr.
\end{equation}
where the domain $U \subset \Rd$ and the boundary condition $f$ are given and assumed to be sufficiently regular.
\begin{center}
	\includegraphics[scale=0.50, trim=0.5cm 3cm 0cm 0.5cm, clip=true]{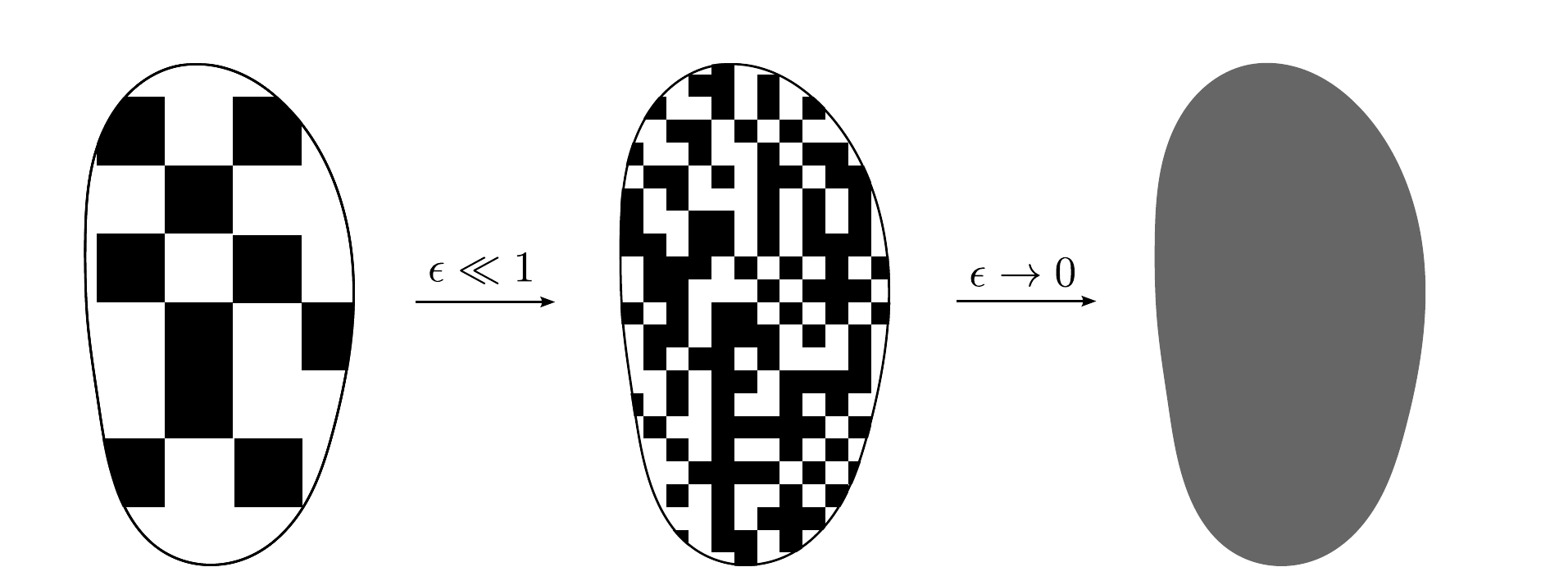}
	\captionof{figure}{As $\ep \to 0$, we expect the medium to ``homogenize'' and that the solution to \eqref{e.pde.eps} approaches the solution of an equation with constant coefficients.}
	\label{f.homog}
\end{center}
As Figure~\ref{f.homog} illustrates, as $\ep$ tends to $0$, it is natural to expect the solution $u_\ep$ to resemble the solution of an effective ``homogenized'' equation with constant coefficients. This is indeed what happens, as has been proved in \cite{K1, Y0, PV1}: there exists a matrix $\ahom$ such that with probability one, the function $u_\ep$ converges in $L^2(U)$ to the solution $\bar u$ of the equation
\begin{equation}  
\label{e.pde.homog}
\Ll\{
\begin{array}{ll}  
-\nabla \cdot \ahom \nabla \bar u = 0  & \quad \text{in } U ,\\
\bar u = f & \quad \text{on } \partial U.
\end{array}
\Rr.
\end{equation}
The crucial feature of this result is that the matrix $\ahom$ is now constant in space, and deterministic. Also, it depends neither on the domain $U$ nor on the boundary condition $f$; it is only a function of the law of the original coefficient field.

\smallskip

As a matter of fact, the structure of elliptic equations allows us a much more direct access to gradients and fluxes of solutions than to the solution itself. Even if one is only interested in the convergence of $u_\ep$ to $\bar u$ stated above, it is thus much more convenient to focus on the convergence of the gradients and fluxes first: we have indeed that, with probability one,
\begin{equation}
\label{e.weak.conv}
\nabla u_\ep \rightharpoonup \nabla \bar u \quad \text{and} \quad \a \Ll( \tfrac \cdot \ep \Rr) \nabla u_\ep \rightharpoonup \ahom \nabla \bar u \qquad (\ep \to 0).
\end{equation}
The precise meaning of the arrow $\rightharpoonup$ is to denote weak convergence in $L^2(U)$. The important point to keep in mind is that $\nabla u_\ep(x)$ will typically \emph{not} be close to $\nabla \bar u(x)$: it is only after one takes a local spatial average of $\nabla u_\ep$ that the result will resemble the local average of $\nabla \bar u$. This information on the weak convergence of $\nabla u_\ep$ to $\nabla \bar u$ is more than enough to recover the convergence of $u_\ep$ to $\bar u$ in $L^2(U)$, see for instance \cite[Lemma~1.8]{AKMbook}.

\smallskip

This result of convergence of $u_\ep$ to $\bar u$ can be thought of as a law of large numbers: there are many ``independent random variables'', each contributing little to the overall phenomenon, which are somehow averaged out and together lead to the emergence of an effective deterministic behavior. However, it is crucial to realize that, contrary to what Figure~\ref{f.homog} may lead us to believe,
\begin{equation}
\label{e.ahom.not}
\ahom \text{ is \emph{not} the average of } \a(x).
\end{equation}
This point is easiest to explain in the case of the coefficient field depicted on the bottom left frame of Figure~\ref{f.coef}, where we assume that $\a(x) = \mathrm{Id}$ in the white region, and $\a(x) = 10^{-5} \, \mathrm{Id}$ in the black region, say. In words, the black region essentially acts as an insulator. We imagine first trying to carry a flux of heat from the top to the bottom part of the square bounding the picture. While it would certainly be easier to do so in the absence of the black lines, their presence is only a minor hindrance, since they are mostly aligned in the vertical direction. The situation changes fundamentally if one wants to propagate a flux of heat from left to right: in this case, the presence of the black lines forces the flux to ``make a lot more detours'', and thus a unit temperature difference between the left and right sides of the square will be associated with a much smaller heat flux. It follows that we do not expect the homogenized matrix to be isotropic: denoting by $e_1$ and $e_2$ the unit vectors in the horizontal and vertical directions respectively, we have argued that $|\ahom e_1| \ll |\ahom e_2|$. In particular, this confirms the statement \eqref{e.ahom.not} that $\ahom$ is not the average of $\a(x)$, since the latter is an isotropic matrix. 

\smallskip

In fact, this argument suggests that there will not be any simple formula for calculating the homogenized matrix $\ahom$ as a function of, for instance, the moments of $\a(x)$. The homogenized matrix must summarize more geometric information on the law of $\a(x)$ that measures, among other things, ``the difficulty of going around obstacles''. (One may point out that a simple formula in fact does exist in dimension $d = 1$, but there is not much question about how to ``go around obstacles'' in one dimension.) The word ``homogenization'', as opposed to ``averaging'', is sometimes used specifically in order to stress that the effective parameters are not obtained by taking the average of the microscopic parameters.

\smallskip

The phenomenon of homogenization has a natural interpretation in terms of stochastic processes. Indeed, the differential operator $-\nabla\cdot \a \nabla$ is the infinitesimal generator of a diffusion process, which we may denote by $(X(t), t \ge 0)$. Seen in this light, the convergence of $u_\ep$ to $\bar u$ is essentially equivalent to the statement that the rescaled process $t \mapsto \ep^{-\frac 1 2} X(\ep^{-1} t)$ converges in law to a Brownian motion with covariance matrix $2\ahom$ as $\ep$ tends to $0$ \cite{PV1}. In other words, this probabilistic interpretation recasts the convergence of $u_\ep$ to $\bar u$ as a central limit theorem for a diffusion in the random environment given by $\a(x)$. Despite the caveat of \eqref{e.ahom.not} and the probabilistic interpretation of homogenization as a central limit theorem, we will continue thinking about the convergence of $u_\ep$ to $\bar u$ as a type of law of large numbers.

\smallskip

The statement of homogenization is interesting for several reasons. For one, it is among the simplest theoretical models for which the large-scale effective model is not obtained by a simple average of the microscopic one. This makes it a very suitable test bed for developing new theoretical ideas and techniques that may also have some validity in other settings involving separations of scales. From a computational perspective, the interest is also evident: while the computation of the solution to \eqref{e.pde.eps} becomes arbitrarily difficult as $\ep$ tends to $0$, the solution of the homogeneous problem \eqref{e.pde.homog} can be computed very rapidly (provided that one can calculate the homogenized matrix $\ahom$ effectively). Depending on the quality of approximation desired, this approximation may or may not be sufficiently precise, and in the latter case, we may want to devise higher-order approximations of the solution $u_\ep$. These higher-order approximations may involve additional deterministic corrections, and also, in the spirit of the central limit theorem, a Gaussian random field whose statistics are captured by a handful of additional effective parameters. 

\smallskip

These considerations motivate the development of a \emph{quantitative} theory of homogenization. That is, we would like to evaluate the speed at which the solution $u_\ep$ to~\eqref{e.pde.eps} converges to the homogenized solution $\bar u$. This endeavor can also be seen as a first step towards the construction of the higher-order approximations alluded to above. 

\smallskip

In analogy with the standard central limit theorem for sums of independent random variables, the fundamental ingredient for obtaining rates of convergence should be encoded in the good mixing properties of the coefficient field: recall that we assume for simplicity that the coefficient field has a unit range of dependence. The difficulty of the problem is that the solution $u_\ep$ to \eqref{e.pde.eps} depends in a very complicated, non-linear and non-local way on the coefficient field $\a(x)$. The question boils down to: how do we ``transfer'' the very good mixing properties of the coefficient field into information on the solution $u_\ep$? The first breakthrough was obtained in \cite{GO1,GO2,GNO,GO3}, where the corrector (introduced below) was shown to be essentially of bounded growth (see also \cite{Y1, NS, vardecay} for earlier sub-optimal results). Inspired by the fundamental insights of \cite{NS2,NS}, one of the main ingredients there is the use of ``non-linear'' concentration inequalities such as the Efron-Stein inequality. Recall that the Efron-Stein inequality states that, if $X = (X_1,\ldots, X_N)$ are independent random variables and $X' = (X_1',\ldots,X_N')$ is an independent copy of $X$, then for every function $F : \R^N \to \R$, we have
\begin{equation*}  
\var \Ll[ F(X) \Rr] \le \frac 12 \sum_{i = 1}^N \E \Ll[ \Ll( F(X_1,X_{i-1},X'_i,X_{i+1}, \ldots,X_N) - F(X) \Rr) ^2 \Rr] .
\end{equation*}
The very useful feature of inequalities such as this one is that they make no assumption whatsoever on the structure of the function $F$ (hence the name ``non-linear''). We will not discuss this approach further here, and focus instead on an alternative approach based on renormalization-type arguments. In this alternative approach, we do not try to study solutions of the problem~\eqref{e.pde.eps} directly, and focus instead on energy-type quantities that display a simpler dependence on the coefficients, at least in the large-scale regime.

%
%
%
%
%
%

\section{Renormalization heuristics for homogenization}

In this section, we describe at a highly heuristic level why renormalization ideas should have a powerful take on the understanding of homogenization. We start by recording the following elementary but fundamental observation. Suppose that the coefficient field has very small oscillations, that is, there exists $\de \ll 1$ such that with probability one, for every $x \in \Rd$,
\begin{equation}
\label{e.small.contrast}
(1-\de)\mathrm{Id} \le \a(x) \le (1+\de)\mathrm{Id}.
\end{equation}
In this case, we have, for some universal constant $C < \infty$,
\begin{equation}
\label{e.almost.av}
\Ll| \ahom - \fint \a \Rr| \le C \de^2,
\end{equation}
where we use the informal notation $\fint \a$ to denote the average of $\a(x)$, that is, with $B_r$ denoting the ball of size $r$ centered at the origin and $|B_r|$ denoting its volume,
\begin{equation*}  
\fint \a := \lim_{r \to \infty} \frac{1}{|B_r|} \int_{B_r} \a(x) \, \d x.
\end{equation*}
Since we assume the law of $\a(x)$ to be invariant under translations by vectors in~$\Zd$, we can also write
\begin{equation*}  
\fint \a = \E \Ll[ \int_{[0,1]^d} \a(x) \, \d x \Rr] .
\end{equation*}
In words, the inequality \eqref{e.almost.av} says that, although \eqref{e.ahom.not} is indeed true and $\ahom$ is not the average of $\a(x)$, in the regime of small oscillations of the coefficients, the average of $\a(x)$ provides an approximation of $\ahom$ that is valid \emph{up to second order} in the size of these oscillations. 

\smallskip

The inequality \eqref{e.almost.av} is an immediate consequence of the classical fact that $\ahom$ is bounded above and below by, respectively, the arithmetic and harmonic means of~$\a(x)$. That is, we have
\begin{equation}  
\label{e.compare.ahom}
\Ll( \fint \a^{-1} \Rr) ^{-1} \le \ahom \le \fint \a,
\end{equation}
as will be explained in the next section around \eqref{e.compare.aU} and \eqref{e.other.compare.aU}. For now, we take \eqref{e.almost.av} for granted and proceed to explain why this should lend support for the possibility of a renormalization-type argument that would uncover the phenomenon of homogenization and its rate of convergence.

\smallskip

Instead of encoding widely separated scales via the introduction of a small parameter $\ep$ used to rescale the coefficient field as in \eqref{e.pde.eps}, we adopt the equivalent and more convenient point of view that the coefficient field $\a(x)$ is fixed once and for all, and that we aim to understand problems that are posed over increasingly large scales. The coefficient field $\a(x)$ gives us the correspondence between gradients and fluxes at the microscopic level, while the homogenized matrix~$\ahom$ encodes the same correspondence, but in the limit of infinitely large scales. The idea of renormalization is that there should be ``something happening inbetween'' these two extremes, a sort of ``progressive homogenization'' that allows us to move smoothly accross scales. That is, for any given length scale $r > 0$, there should exist a coefficient field~$\a_r(x)$ encoding the correspondence between gradients and fluxes of solutions ``after we forget about small-scale details below scale $r$''. Ideally, we would then have a ``renormalization map'' telling  us how to calculate the coarser coefficient field, say $\a_{2r}(x)$, from the finer information encoded in $\a_r(x)$. An analysis of this renormalization map should then reveal the rate of convergence to the limit.

\smallskip

Assuming that the identification of such ``coarsened coefficients'' $\a_r(x)$ is indeed possible, we can intuitively guess how they will behave using \eqref{e.small.contrast}-\eqref{e.almost.av}. Indeed, as the length scale $r$ increases to infinity, we expect $\a_r(x)$ to become more and more sharply concentrated around its constant limit $\ahom$. In particular, this procedure of progressive homogenization automatically brings the coarsened coefficient field~$\a_r(x)$ into a regime of small fluctuations very similar to that assumed in \eqref{e.small.contrast}. The relation~\eqref{e.almost.av} expresses a relationship between the ``infinite-scale'' coarsened coefficient field~$\ahom$ and the global average of the microscopic coefficient field $\a(x)$. By analogy, we expect that as $r$ becomes very large, the approximation of $\a_{2r}(x)$ by a local average of $\a_r(x)$ should become asymptotically sharp, up to a lower order error. If this is so and this procedure of progressive homogenization indeed becomes closer and closer to a simple local averaging procedure as $r \to \infty$, then we should expect $\a_r(x)$ to have asymptotic fluctuations dictated by the scaling of the central limit theorem. That is, we expect $\a_r(x) - \ahom$ to have Gaussian fluctuations of typical size $r^{-\frac d 2}$.

\smallskip

Continuing with intuitive arguments, we explore possible consequences of this prediction (and along the way hopefully start to clarify, still on an informal level, a somewhat more precise meaning for the coarsened coefficient field $\a_r(x)$). For simplicity, let us fix $p \in \Rd$ and consider solving for the Dirichlet problem on a very large cube $\cu$ with affine boundary condition $x \mapsto p\cdot x$ on $\partial \cu$. Clearly, the solution to the corresponding homogenized problem is simply the affine function $x \mapsto p\cdot x$. In order to emphasize this, we write down the solution to the heterogeneous problem in the form $x \mapsto p\cdot x + \phi_p(x)$. That is, we solve for
\begin{equation}  
\label{e.pde.corrector}
\Ll\{
\begin{array}{ll}  
-\nabla \cdot \a(p+\nabla \phi_p) = 0  & \quad \text{in } \cu ,\\
\phi_p = 0 & \quad \text{on } \partial \cu,
\end{array}
\Rr.
\end{equation}
and we are particularly interested in showing that $|\phi_p(x)| \ll |x|$ for large values of~$|x|$. Although we will not justify this, it is possible to make sense of $\nabla \phi_p$ also in the limit where the size of the cube $\cu$ is sent to infinity, and this limiting object is called the gradient of the \emph{corrector} in the direction of $p$. (Contrarily to its gradient, the function~$\phi_p$ itself may fluctuate unboundedly as the cube is blown up to infinity. In fact, its behavior can be compared with that of a Gaussian free field\footnote{The name ``Gaussian free field'' has become standard in the probability literature, although it is a bit redundant, since a random field is said to be free if it is Gaussian. The name ``masless free field'' is usually favored in the theoretical physics literature.}  that is regularized on a unit scale, as will be seen below.)

\smallskip

As was explained around \eqref{e.weak.conv}, although we may be mostly interested in obtaining quantitative estimates on the growth of $\phi_p$ itself, we will have more direct access to information about its gradient and flux. We thus aim instead to give quantitative estimates on their weak convergence. More precisely, we would like to know the rate at which the spatial average of $\nabla \phi_p$ over a region of size $r$ converges to $0$; and similarly for the convergence of spatial averages of $\a(p + \nabla \phi_p)$ towards $\ahom p$. 

\smallskip

We aim to obtain this information from our prediction on the asymptotic size of the difference $\a_r(x) - \ahom$. In order to do so, we need to ``forget about small scale details'' of the solution $x \mapsto p\cdot x + \phi_p(x)$ below scale $r$. For instance, we may take a smooth function $\chi \in C^\infty_c(\Rd)$ with compact support and satisfying $\int_\Rd \chi = 1$, and then define, for every $r \ge 1$, the convolution
\begin{equation*}  
\phi_{p,r} := \phi_p \ast \Ll(r^{-d} \chi \Ll( \tfrac{\cdot}{r} \Rr)\Rr) .
\end{equation*}
The fundamental idea of the coarsened coefficient field $\a_r(x)$ is that it should govern, at least approximately, the behavior of this coarsened solution $\phi_{p,r}$, in the sense that 
\begin{equation*}  
-\nabla \cdot \a_r \Ll( p + \nabla \phi_{p,r} \Rr) \simeq 0.
\end{equation*}
Denoting
\begin{equation*}  
\mathsf W_r(x) := \a_r(x) - \ahom,
\end{equation*}
we may rearrange the approximate equation above in the form
\begin{equation}  
\label{e.coarse.eq}
-\nabla \cdot \a_r \nabla \phi_{p,r} \simeq \nabla \cdot \Ll( \msf W_r p \Rr) .
\end{equation}
Recalling our prediction that the size of $\msf W_r(x)$ should be of the order of $r^{-\frac d 2}$, a classical testing of the equation above against $\phi_{p,r}$ suggests that
\begin{equation}  
\label{e.size.phipr}
\Ll|\nabla \phi_{p,r}\Rr| \ls r^{-\frac d 2}.
\end{equation}
This is exactly the information we were looking for: indeed, since gradients and convolutions commute, the estimate above states that the average of $\nabla \phi_p$ over a region of size $r$ decays to $0$ like $r^{-\frac d 2}$. Once such information is obtained, one can recover sharp information on the growth of $\phi_p$ itself: it grows like $\log^\frac 1 2|x|$ in dimension $d = 2$, and remains essentially bounded in dimension $d \ge 3$ (see \cite[Theorem~4.1]{AKMbook}).

\smallskip

\begin{figure}
\centering
\begin{tabular}{cc}
\includegraphics[width = 0.45\textwidth]{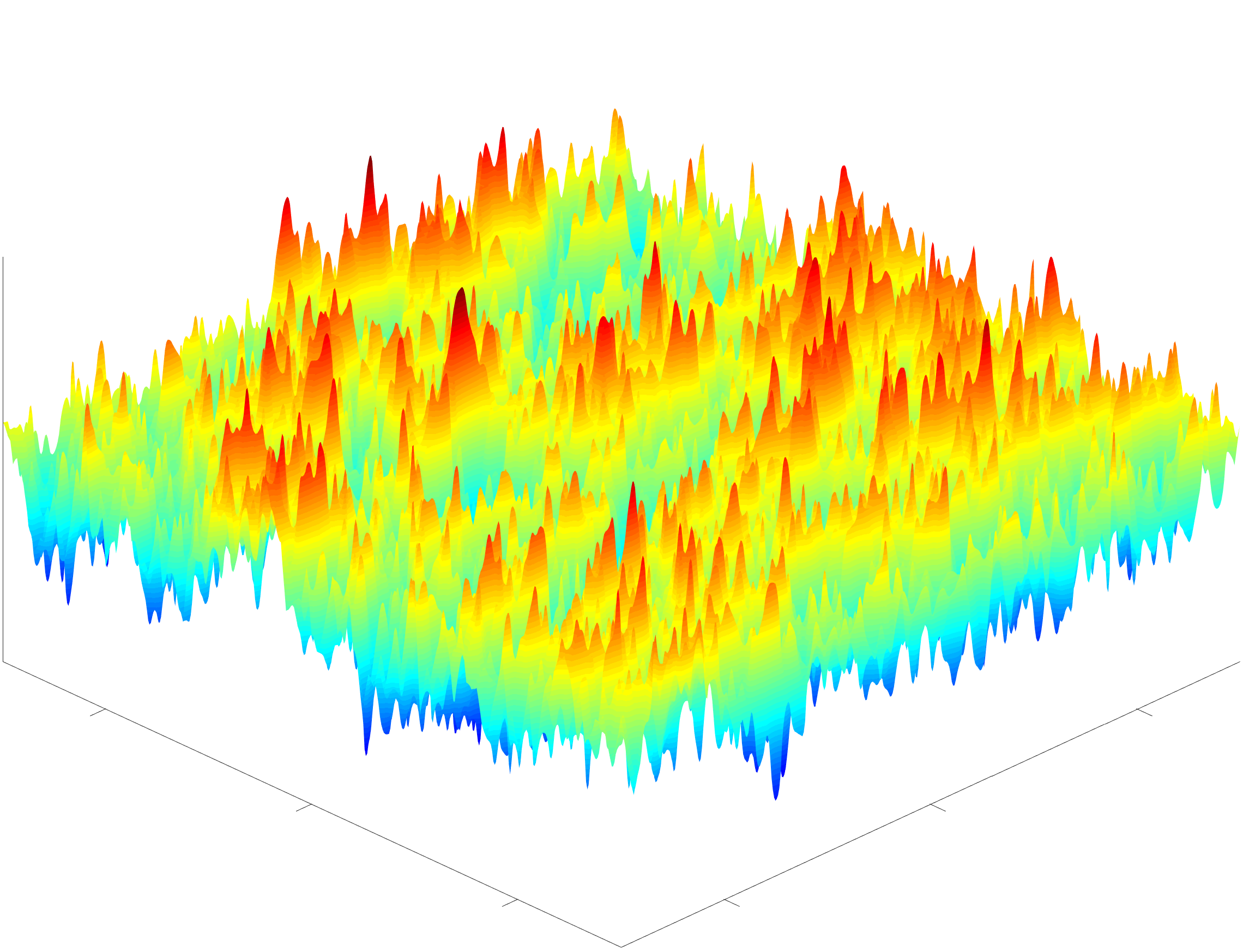}

&

\includegraphics[width = 0.45\textwidth]{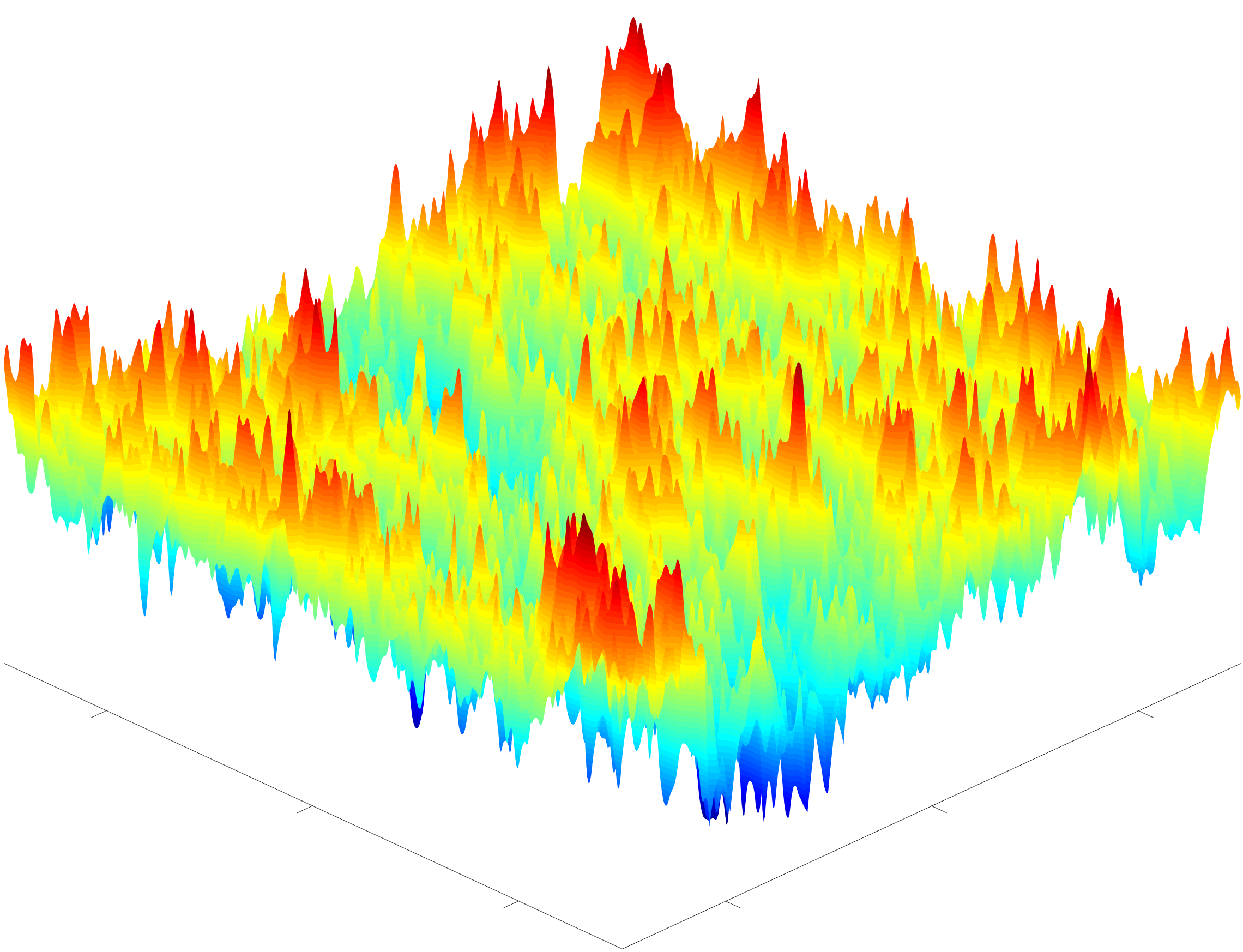} 

 \vspace{0.5cm}
 \\

\includegraphics[width = 0.45\textwidth]{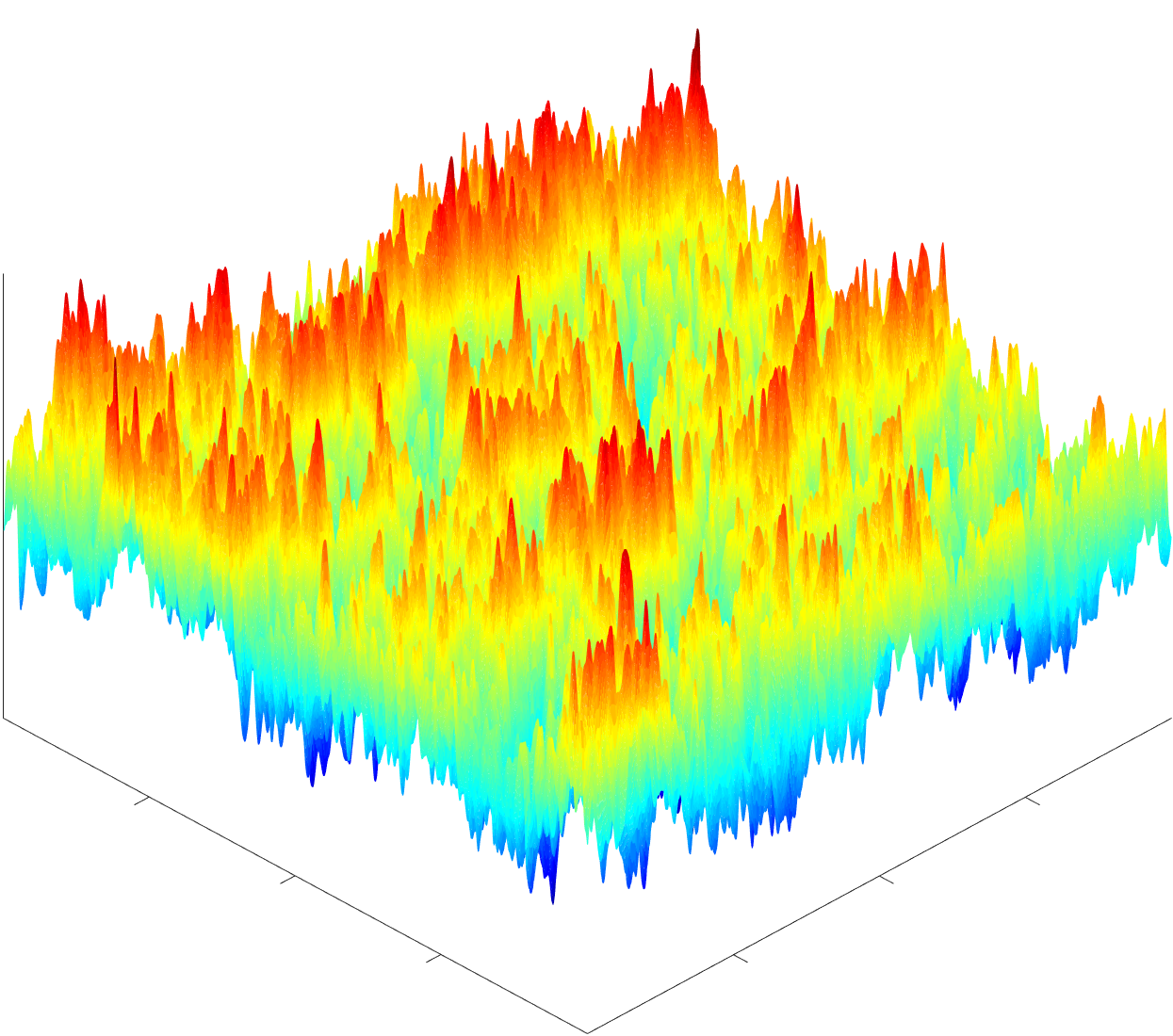}

& 

\includegraphics[width = 0.45\textwidth]{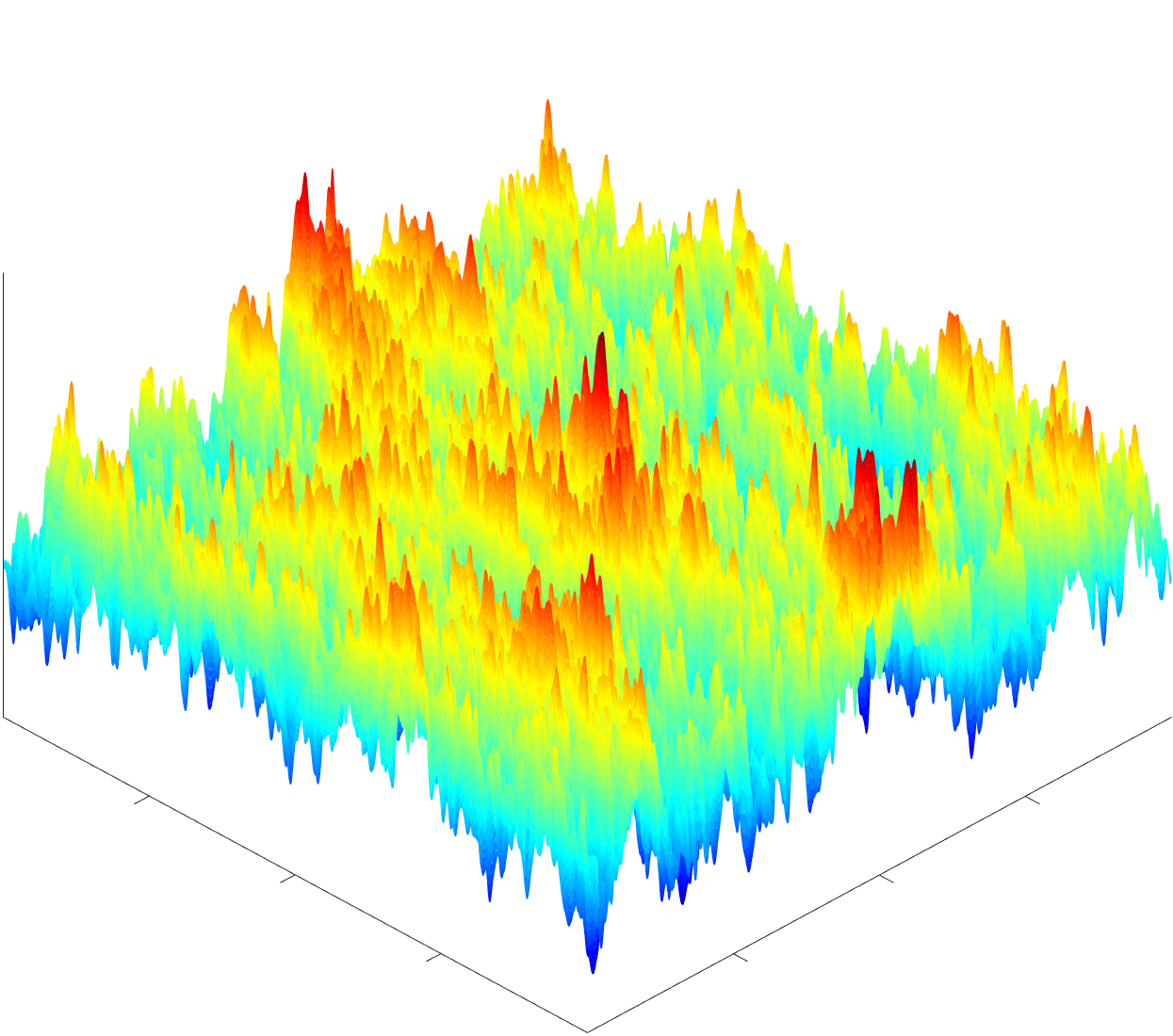}
\end{tabular}
\caption{Top: correctors $\phi_{e_1}$ and $\phi_{e_2}$ in the directions of the two basis vectors~$e_1$ and $e_2$, for random coefficients as in the top left frame of Figure~\ref{f.coef}. Bottom: examples of Gaussian random fields with the structure predicted by \eqref{e.approx.gff}, after convolution with a smooth bump function on a unit scale.}
	\label{f.gff}
\end{figure}

In fact, even finer information can be read off from the approximate equation \eqref{e.coarse.eq}. Indeed, since the prediction for the size of $\msf W_r$ was obtained via a central limit theorem argument, the reasoning in fact yields a more refined description of this quantity: namely, we expect it to closely resemble the convolution on scale~$r$ of a white noise random field. Moreover, in view of \eqref{e.size.phipr} and of the fact that $|\a_r - \ahom| \ls r^{-\frac d 2}$, we see that replacing $\a_r$ by $\ahom$ in \eqref{e.coarse.eq} produces a lower-order error, and thus
\begin{equation}  
\label{e.approx.gff}
-\nabla \cdot \ahom \nabla \phi_{p,r} \simeq \nabla \cdot \Ll( \msf W_r p \Rr) .
\end{equation}
Replacing $\simeq$ by $=$ and $\msf W_r$ by a true white noise, we would obtain the equation defining a Gaussian random field very similar to a Gaussian free field (see \cite[Section~5.1]{AKMbook}). The approximate equation \eqref{e.approx.gff} thus suggests that the law of the large-scale convolution of $\phi_p$ becomes asymptotically very close to the law of the large-scale convolution of this Gaussian random field, a result that can be compared with a central limit theorem; see \cite[Theorem~5.24]{AKMbook} for a precise statement, and Figure~\ref{f.gff} for an illustration.

%
%
%
%
%
%

\section{A mathematical approach to renormalization for homogenization}

In my opinion, the intuitive picture described in the previous section is extremely compelling, and strongly calls for a mathematical embodiement. It is the purpose of the present section to review, at a more precise but still highly informal level, the outline of such an approach as developed in \cite{AS,AM,AKM,AKM2}; see also \cite{GO5} for an alternative approach in a similar spirit, and \cite{AKMbook} for a book on the topic.

\smallskip

The most fundamental task is to identify a candidate for the notion of coarsened coefficients discussed in the previous section. Borrowing an idea of \cite{DM1,DM2}, we introduce, for each bounded domain $U \subset \Rd$ and $p \in \Rd$, the energy-type quantity
\begin{equation}
\label{e.def.nu}
\nu(U,p) := \inf_{u \in \ell_p + H^1_0(U)} \fint_U \frac 1 2 \nabla v \cdot \a \nabla v,
\end{equation}
where $\ell_p(x) := p\cdot x$, we use the notation $\fint_U := |U|^{-1} \int_U$ for the integral normalized by the volume of $U$, and $H^1_0(U)$ denotes the set of functions  in $L^2(U)$ with gradient in $L^2(U)$ and which vanish on $\partial U$. It is classical to verify that the minimizer of this problem exists and is unique, and is simply the $\a$-harmonic function with boundary condition given by $p\cdot x$. (We say that a function $v$ is $\a$-harmonic if it satisfies $-\nabla \cdot \a \nabla v = 0$.)

\smallskip

The quantity $\nu(U,p)$ satisfies a number of remarkable properties. In particular, the mapping $U \mapsto \nu(U,p)$ is subadditive. In order to verify this, consider a partition of the domain $U$ into subdomains $U_1,\ldots,U_k$. Then the minimizers $v_1,\ldots,v_k$, defined on $U_1,\ldots,U_k$ respectively, share a consistent boundary condition on the interfaces between subdomains. As a consequence, we can glue them together and obtain a candidate function for the minimization in the larger domain $U$. It thus follows that 
\begin{equation}  
\label{e.subadd}
\nu(U,p) \le \sum_{i = 1}^k \frac{|U_i|}{|U|} \nu(U_i,p).
\end{equation}
Another elementary but very useful property of $\nu(U,p)$ is that it depends only on the coefficient field inside $U$. Finally, for convenience, we also note that the mapping $p \mapsto \nu(U,p)$ is quadratic. This follows from the fact that the solution of the Dirichlet problem with boundary condition $p\cdot x$ depends linearly on $p$. Hence, there exists a symmetric matrix $\a(U)$ such that for every $p \in \Rd$,
\begin{equation} 
\label{e.def.aU}
\nu(U,p) = \frac 1 2 p\cdot \a(U) p.
\end{equation}
This matrix $\a(U)$ is our candidate ``homogenized matrix on the scale of the domain~$U$''. It encodes the (spatially averaged) energy of solutions of Dirichlet problems on the domain $U$ with affine boundary condition. For every $r > 0$, we denote the cube of side length $r$ centered at the origin by
\begin{equation*}  
\cu_r := \Ll( -\frac r 2, \frac r 2 \Rr) ^d.
\end{equation*}
The subadditivity property \eqref{e.subadd} implies that with probability one, the quantity $\nu(\cu_r,p)$ converges to some limit as $r$ tends to infinity. This can be obtained by an application of the very general subadditive ergodic theorem \cite{AK}, or more directly using the stronger independence assumptions at our disposal, see \cite[Proposition~1.3]{AKMbook}. Since the mapping $p \mapsto \nu(\cu_r,p)$ is quadratic, the limit will also be quadratic, and we \emph{define} the matrix $\ahom$ so that
\begin{equation}  
\label{e.lim.nu}
\lim_{r \to \infty} \nu(\cu_r,p) = \frac 1 2 p \cdot \ahom p.
\end{equation}
It is then our task to show that homogenization happens with this definition of the matrix $\ahom$, and to quantify the rate of convergence. 

\smallskip

The fact that $\a(U)$ is defined as the spatially averaged energy of certain solutions is a first indication that it may serve as a suitable notion of coarsened coefficient field. We can also witness that this matrix also encodes a correspondence between coarsened versions of the gradients and fluxes of solutions. Indeed, denote by $v(\cdot,U,p)$ the minimizer in the definition \eqref{e.def.nu} of $\nu(U,p)$. Since $v \in \ell_p + H^1_0(U)$, we have
\begin{equation}
\label{e.spat.av.grad}
\fint_U \nabla v(\cdot,U,p) = p.
\end{equation}
Moreover, for every $p,p' \in \Rd$, we have
\begin{equation*}  
p'\cdot \a(U) p = \fint_U \nabla v(\cdot,U,p') \cdot \a \nabla v(\cdot,U,p).
\end{equation*}
Indeed, this identity with $p' = p$ boils down to definitions; the more general version then follows using elementary symmetry properties of the bilinear forms. Recalling that $v(\cdot,U,p') \in \ell_{p'} + H^1_0(U)$, and that $v(\cdot, U,p)$ is an $\a$-harmonic function, we deduce that 
\begin{equation*}  
\fint_U \nabla \Ll( v(\cdot,U,p') - \ell_{p'} \Rr) \cdot \a \nabla v(\cdot,U,p) = 0,
\end{equation*}
and thus
\begin{equation*}  
p'\cdot \a(U) p = \fint_U p' \cdot \a \nabla v(\cdot,U,p),
\end{equation*}
that is,
\begin{equation*}  
\a(U) p = \fint_U \a \nabla v(\cdot, U,p).
\end{equation*}
Comparing this last identity with \eqref{e.spat.av.grad}, we thus obtain that $\a(U)$ transfers information about the spatial average of the gradient of a solution into information about the spatial average of its flux.

\smallskip

From the definition of $\ahom$ in \eqref{e.lim.nu}, we can already justify the rightmost inequality appearing in \eqref{e.compare.ahom}. Indeed, by choosing the affine function $\ell_p$ as a candidate minimizer in \eqref{e.def.nu}, we immediately get that
\begin{equation}  
\label{e.compare.aU}
\a(U) \le \fint_U \a.
\end{equation}
The desired inequality between $\ahom$ and the average of $\a(x)$ follows by choosing~$U = \cu_r$ and then letting $r$ tend to infinity. For the leftmost inequality in \eqref{e.compare.ahom}, one can first show that 
\begin{equation*}  
\nu(U,p) = \sup_{\g} \fint_U \Ll( -\frac 1 2 \g \cdot \a^{-1} \g + p \cdot \g \Rr) ,
\end{equation*}
where the supremum is taken over every divergence-free vector field $\g \in L^2(U)$ (see \cite[(2.8)]{AKMbook}). This is a ``dual'' formulation of the Dirichlet problem that puts emphasis on fluxes rather than gradients of solutions. Choosing $\g = q$ for a fixed $q \in \Rd$ (constant flux instead of constant gradient as in the previous argument) then yields that 
\begin{equation}  
\label{e.baby.dual}
\nu(U,p) \ge p\cdot q - \frac 1 2 q\cdot\Ll(\fint_U  \a^{-1}\Rr) q.
\end{equation}
Optimizing over $q$, we obtain that
\begin{equation}  
\label{e.other.compare.aU}
\nu(U,p) \ge \frac 1 2 p \cdot \Ll( \fint_U \a^{-1} \Rr) ^{-1} p,
\end{equation}
and this implies the leftmost inequality in \eqref{e.compare.ahom}.

\smallskip

This reasoning suggests the possible relevance to study not only solutions of Dirichlet problems with affine boundary conditions, but also Neumann problems. We thus introduce a corresponding energy: for every bounded domain $U$ and $q \in \Rd$, we set
\begin{equation*}  
\nu^*(U,q) := \sup_{u \in H^1(U)} \fint_U \Ll( - \frac 1 2 \nabla u \cdot \a \nabla u + q \cdot \nabla u\Rr) .
\end{equation*}
The maximizer of this problem is unique up to the addition of a constant, and is the $\a$-harmonic function such that $(\a\nabla u - q)\cdot \mathbf n$ vanishes on $\partial U$, where $\mathbf n$ denotes the outer normal to $\partial U$. The mapping $U \mapsto \nu^*(U,q)$ is also subadditive, for reasons ``opposite'' to $\nu$: the maximizer in the definition of $\nu^*(U,q)$ restricted to subdomains gives us candidate functions on the subdomains.

\smallskip

Recalling \eqref{e.spat.av.grad} and using the minimizer $v(\cdot,U,p)$ of $\nu(U,p)$ as a candidate for this problem, we obtain that for every $p,q \in \Rd$,
\begin{equation}  
\label{e.convex.dual}
\nu^*(U,q) \ge p\cdot q - \nu(U,p).
\end{equation}
Notice the similarity between this inequality and that in \eqref{e.baby.dual}. 
In words, the mapping $q \mapsto \nu^*(U,q)$ is an upper bound for the convex dual of the mapping $p \mapsto \nu(U,p)$. Since we expect homogenization to happen in the large-scale limit, it is plausible that over large scales, the maximizer in the definition of $\nu^*(U,q)$ becomes very close to some function $v(\cdot,U,p)$ for the appropriate choice of the slope $p$. That is, we expect $\nu(U,\cdot)$ and $\nu^*(U,\cdot)$ to become approximately convex dual to one another in the large-scale limit. This suggests in particular that
\begin{equation}
\label{e.lim.nu*}
\lim_{r \to \infty} \nu^*(\cu_r,q) = \frac 1 2 q \cdot \ahom^{-1} q.
\end{equation}

\smallskip

The rough outline allowing to obtain the predicted rate of convergence in the limit \eqref{e.lim.nu} can be described as follows.

\begin{enumerate}  
\item We show the existence of a possibly very small exponent $\alpha > 0$ such that
\begin{equation}
\label{e.smallalpha}
\Ll| \nu(\cu,p) - \frac 1 2 p \cdot \ahom p \Rr| \ls |\cu|^{-\al}.
\end{equation}

\item By \eqref{e.smallalpha}, the coarsened coefficients on the scale of a given cube $\cu$ fluctuate by approximately $|\cu|^{-\al}$. In view of \eqref{e.small.contrast}-\eqref{e.almost.av}, we thus expect that 
\begin{equation}  
\label{e.additive}
\Ll| \nu(\cu,p) - 2^{-d} \sum_{z \in \mcl Z} \nu(z + \cu',p) \Rr| \ls |\cu|^{-2\al},
\end{equation}
where we decomposed the cube $\cu$ into $2^d$ subcubes of half the side length, which we denote by $(z + \cu')_{z \in \mcl Z}$.

\item By \eqref{e.additive}, the quantity $\nu$ is almost additive, and thus, using independence,
\begin{equation}
\label{e.indep}
\Ll|  \nu(\cu,p) - \E \Ll[ \nu(\cu,p) \Rr] \Rr| \ls |\cu|^{-\Ll((2\al) \wedge \frac 1 2\Rr)}.
\end{equation}
\end{enumerate}
In \eqref{e.indep}, we use the notation $(2\al) \wedge \frac 1 2$ to denote the minimum between $2\al$ and $\frac 1 2$. 
From \eqref{e.additive}, we also infer that the difference between the expectation of $\nu(\cu,p)$ and its limit $\frac 1 2 p \cdot \ahom p$ is bounded by $C|\cu|^{-2\al}$. In this rough outline, we thus end up with the conclusion that we can in fact replace the exponent $\al$ in \eqref{e.smallalpha} by the exponent $(2\al) \wedge \frac 1 2$. In principle, the reasoning could then be iterated until all statements are known with $\alpha$ replaced by the exponent $\frac 1 2$, in agreement with the prediction of the previous section.

\smallskip

In fact, we already know from \eqref{e.subadd} that the quantity between absolute values in \eqref{e.additive} is nonpositive. The control of this quantity from the other side should resemble the proof of the leftmost inequality in \eqref{e.compare.ahom}, but at the level of coarsened coefficients. The quantity $\nu^*(U,q)$ is a fundamental tool for this purpose. 

\smallskip

Part (3) of the outline above is the simplest to understand. Indeed, if we pretend for a moment that the additivity of $U\mapsto \nu(U,p)$ is exact (that is, that the left side of \eqref{e.additive} is actually zero), then the calculation of $\nu(\cu,p)$ really boils down to taking the average of the cubes at a slightly smaller scale. We can thus apply classical techniques to prove central limit theorems for sums of independent random variables, and obtain fluctuations of the order of $|\cu|^{-\frac 1 2}$. In practice, the result can only be as good as the error in the additivity statement \eqref{e.additive}, and this explains the exponent on the right side of \eqref{e.indep}.

\smallskip

We now say a few words on how to obtain the statement \eqref{e.smallalpha} for some possibly very small exponent $\al > 0$. The real difficulty is to control the rate of convergence of the expectation of $\nu(\cu_r,p)$; fluctuations can then be recovered using subadditivity (to be precise, subadditivity of $\nu$ and $\nu^*$ allow to control their respective upper fluctuations, and we can relate the lower fluctuations of $\nu$ with the upper fluctuations of $\nu^*$). In order to obtain this small exponent in the rate of convergence of the quantity $\E \Ll[ \nu(\cu_r,p) \Rr]$ towards its limit, it would be sufficient to assert the existence of a constant $C < \infty$ such that for every $r \ge 1$,
\begin{equation}  
\label{e.contract}
\E \Ll[  \nu(\cu_r,p) \Rr] - \frac 1 2 p \cdot \ahom p \le C \Ll(\E \Ll[  \nu(\cu_r,p) \Rr]  - \E \Ll[  \nu(\cu_{2r},p) \Rr]\Rr).
\end{equation}
(Subadditivity and stationarity imply that both sides are nonnegative). Indeed, a rearrangement would then give that
\begin{equation*}  
\E \Ll[  \nu(\cu_{2r},p) \Rr]- \frac 1 2 p \cdot \ahom p \le \frac{C}{C-1} \Ll( \E \Ll[  \nu(\cu_{r},p) \Rr]- \frac 1 2 p \cdot \ahom p \Rr) ,
\end{equation*}
which could be iterated to produce the desired decay in $r$, at least along powers of~$2$. It seems however very difficult to show an inequality such as \eqref{e.contract} directly, in particular because $\ahom$ is defined as an infinite-volume limit, while we have more direct access to finite-volume quantities. The idea of \cite{AS} is to try instead to prove a ``finite-volume'' statement of the form
\begin{equation}  
\label{e.better.contract}
\E \Ll[ \nu(\cu_r,p) \Rr] - \sup_{q \in \Rd} \Ll( p\cdot q - \E\Ll[\nu^*(\cu_r,q) \Rr] \Rr) \le C \tau(r),
\end{equation}
where similarly to the right side of \eqref{e.contract},  the quantity $\tau(r)$ is meant to capture the ``additivity defect'' for the quantities $\nu$ and $\nu^*$, for instance
\begin{equation*}  
\tau(r) := \sup_{|p| \le 1} \Ll( \E \Ll[  \nu(\cu_r,p) \Rr]  - \E \Ll[  \nu(\cu_{2r},p) \Rr] \Rr)  + \sup_{|q| \le 1} \Ll( \E \Ll[  \nu^*(\cu_r,q) \Rr]  - \E \Ll[  \nu^*(\cu_{2r},q) \Rr] \Rr).
\end{equation*}
In view of \eqref{e.lim.nu*}, we expect the supremum over $q$ on the left side of \eqref{e.better.contract} to approach $\frac 1 2 p \cdot \ahom p$ as $r$ tends to infinity. In words, the inequality \eqref{e.better.contract} states that the ``convex duality defect'' between $\nu$ and $\nu^*$, see \eqref{e.convex.dual}, can be controlled by the additivity defect between two scales. Let us argue very briefly on the plausibility of the idea that a small additivity defect should imply a small convex duality defect. Subadditivity of $\nu$ and $\nu^*$ is obtained by comparing the optimizer on the larger scale with a patching of essentially independent and identically distributed pieces on the smaller scale. The smallness of the additivity defect should indicate that these two objects are relatively close. This in turn should imply that the large-scale maximizer remains relatively close to an affine function. Finally, if we know that the maximizer for $\nu^*(\cu_r,q)$ is close to an affine function of slope $p$, say, then we would expect it to be close to the function $v(\cdot, \cu_r,p)$ as well (indeed, if $v(\cdot,\cu_r,p)$ is close to some affine function, then this must be the affine function with slope~$p$, by~\eqref{e.spat.av.grad}). From this, we would then deduce that the inequality \eqref{e.convex.dual} is in fact almost an equality, as desired. A more accurate discussion would involve quantifying closeness to affine functions in $L^2(\cu_r)$, and then using the Caccioppoli inequality to deduce closeness of the optimizers in $H^1(\cu_r)$. Precise statements and proofs can be found in \cite[Chapter~2]{AKMbook}. 

\smallskip

We now discuss the meaning of the symbols $\ls$ in \eqref{e.smallalpha}-\eqref{e.indep}. They are not only meant to hide multiplicative constants. Indeed, in any finite volume, there is some chance for the coefficient field to be completely atypical, and in this scenario, the left side of \eqref{e.smallalpha} will \emph{not} be small. The inequalities \eqref{e.smallalpha}-\eqref{e.indep} should thus be interpreted as statements that hold with high probability. That is, for any random variable $X$ and $\theta > 0$, the notation $X \lesssim \theta$ should encode that the random variable $\theta^{-1} X$ remains tightly controlled. Since we appeal to the central limit theorem in the derivation of \eqref{e.indep}, a good notion for what ``tightly controlled'' could mean is that the random variable has Gaussian tails. More generally, for any $s > 0$, we use the notation 
\begin{equation}
\label{e.notation}
X \le \O_s(\theta) \quad \iff \quad \E \Ll[ \exp \Ll( (\theta^{-1} X)_+^s \Rr) \Rr] \le 2,
\end{equation}
where $x_+ := \max(x,0)$ denotes the positive part of $x \in \R$. The case $s = 2$ corresponds to a Gaussian tail behavior, see \cite[Appendix~A]{AKMbook}. This notation helps to clarify an important difficulty in the inductive argument sketched above: if we have \eqref{e.smallalpha} with right side of the form $\O_s \Ll( C |\cu|^{-\al}\Rr)$, then the reasoning in step (2) leads to an inequality \eqref{e.additive} with right side of the form $\Ll( \O_s(C|\cu|^{-\al} )\Rr) ^2$, that is, $\O_{s/2} \Ll( C^2 |\cu|^{-2\al} \Rr)$. While the size of the error itself has improved, the exponent $s$ of stochastic integrability has been divided by a factor of $2$. In practice, this means that the probability of large deviations is no longer estimated sharply. We circumvent this difficulty by always relying on our initial estimate \eqref{e.smallalpha} with very small exponent $\alpha$, for which we briefly sketched the argument above, in order to estimate these large deviation events. Technically, it then becomes more convenient to simply improve $\alpha$ by a small fixed amount at each step of the iteration, with the ``moderate deviations'' being improved by squaring, while keeping the seed ``large deviation'' estimate intact from the onset. As a consequence, it is fundamental that the initial estimate \eqref{e.smallalpha} with very small exponent $\al > 0$ gives us very strong control on the probability of rare events. The initial estimate should thus give us in particular a statement of the following form: that for any fixed $\ep > 0$, one can find $\al > 0$ and a constant $C< \infty$ such that for every cube $\cu$,
\begin{equation*}  
\P \Ll[ \Ll| \nu(\cu,p) - \frac 1 2 p \cdot \ahom p \Rr| \le |\cu|^{-\al} \Rr] \le C \exp\Ll(-|\cu|^{(1-\ep)}\Rr).
\end{equation*}
It is easy to see that this estimate is essentially best possible. Indeed, taking the ``random cherckerboard'' depicted in the top left frame of Figure~\ref{f.coef} as an example of coefficient field, we see that the probability to observe only white unit squares in a given large cube $\cu$ is of the order of $\exp(-c|\cu|)$, and in this case, the difference $\nu(\cu,p) - \frac 1 2 p\cdot \ahom p$ will be of order one.

\smallskip

The final, and technically most problematic, caveat to the outline described in the steps (1-3) above is that in fact, the optimizers for $\nu$ and $\nu^*$ cannot be close to one another at very high precision, since they will be at a distance of order $1$ from one another at least in a unit-scale neighborhood of the boundary. The presence of this boundary layer means that it is not possible to establish the inequality \eqref{e.additive} for a right side which is smaller than $|\cu|^{-\frac 1 d}$, a suboptimal conclusion that was the main result of \cite{AKM}. Going past this boundary phenomenon requires to redefine the quantities $\nu$ and $\nu^*$ in a ``smoother way'', and was the main achievement of \cite{AKM2}. The whole strategy is exposed in details in \cite[Chapter~4]{AKMbook}.

\smallskip

A core ingredient entering the formalization of this strategy, first introduced for this purpose in \cite{AS}, is the notion of large-scale regularity of $\a$-harmonic functions. It is simplest to explain this idea by analogy with the more classical Schauder estimates. These estimates state that if the coefficient field $\a$ is $\alpha$-H\"older continuous, then any $\a$-harmonic function~$u$ in the unit ball $B_1$ is Lipschitz continuous (in fact, $C^{1,\al}$) in the interior of the ball---see \cite[(3.8)]{AKMbook} for a quantitative statement. The idea of the proof is to compare, over a series of balls with geometrically decreasing radius, the true solution with the solution of the equation with ``frozen'', constant coefficients given by the value of the true coefficient field at the center of the ball. This allows to transfer the higher regularity properties of solutions of equations with constant coefficients onto the solution of the heterogeneous equation; see \cite[(3.9)-(3.10)]{AKMbook} and \cite[Chapter~3]{HL} for a more precise discussion. 

\smallskip

For equations with random coefficients, we use a similar idea, except that it is over the \emph{largest} scales that $\a$-harmonic functions become closer and closer to solutions of constant-coefficient equations. That is, once the validity of \eqref{e.smallalpha} for a small exponent $\al > 0$ is established, one can deduce an error estimate for homogenization which one then uses to replace the argument involving a local ``freezing'' of the coefficients in the classical Schauder estimate. The conclusion of the argument is, roughly speaking, that $\a$-harmonic functions are essentially ``Lipschitz continuous from the largest scale down to the unit scale''. (Naturally, homogenization does not inform us about regularity on scales smaller than the unit scale.) That is, for every $r \ge 1$ and every $\a$-harmonic function $u$ in the ball $B_r$, we have 
\begin{equation*}  
\fint_{B_1} |\nabla u|^2 \ls \fint_{B_r} |\nabla u|^2,
\end{equation*}
where the implicit constant is independent of the radius $r$ (and $\ls$ still hides probabilistic quantifiers). Using the Caccioppoli inequality, see \cite[Lemma~C.2]{AKMbook}, and invoking the result above with $r$ replaced by $r/2$, we obtain that, for every $r \ge 1$,
\begin{equation}
\label{e.regularity}
\fint_{B_1} |\nabla u|^2 \ls r^{-1} \fint_{B_r} u^2.
\end{equation}
As a striking illustration of the power of this estimate, we now observe that it immediately implies gradient estimates for Green functions. 
Restricting ourselves to dimensions $d \ge 3$ for simplicity, we denote by $(G(x,y),x,y \in \Rd)$ the Green function associated with the operator $-\nabla \cdot \a \nabla$ on $\Rd$. It is well-known (and can be deduced from the similar parabolic construction recalled in \cite[Appendix~E]{AKMbook}) that for arbitrary coefficient fields satisfying the uniform ellipticity condition \eqref{e.unif.ellip}, the Green function exists and satisfies the bound
\begin{equation*}  
G(x,y) \le \frac{C}{|x-y|^{d-2}}.
\end{equation*}
Since the function $x \mapsto G(x,0)$ is $\a$-harmonic in the ball $B_{|x|/2}(x)$ of radius $\frac{|x|}{2}$ centered at $x$, we can apply the estimate \eqref{e.regularity} in this ball and conclude that 
\begin{equation*}  
\Ll(\fint_{B_1(x)} |\nabla_x G(\cdot,0)|^2\Rr)^\frac 1 2 \ls \frac C {|x|^{d-1}}.
\end{equation*}
The precise formulation and proof of the regularity estimate \eqref{e.regularity}, as well as higher-order versions thereof, are explained in \cite[Chapter~3]{AKMbook}. The consequences of these estimates for parabolic and elliptic Green functions (in any dimension) are worked out in \cite[Chapter~8]{AKMbook}, and more refined information on the homogenization of the Green function is obtained in \cite[Section~9.2]{AKMbook}. The latter results may be of particular interest to probabilists, since they can be interpreted as a quantitative quenched local central limit theorem for the diffusion process.

\smallskip

To sum up, one can implement a rigorous version of the ideas exposed in the previous section and in the rough outline above, subject to some caveats, the most important of which being the problem caused by boundary layers. Once this is achieved and a sharp estimate on the fluctuations of the coarsened coefficients is known, one can essentially reproduce the argument leading to \eqref{e.size.phipr} and obtain optimal estimates on the large-scale behavior of the correctors $\phi_p$. This can be considered as the most fundamental result of \cite{AKMbook}, and is achieved in Chapter~4 therein. Following the outline leading to \eqref{e.approx.gff}, one can refine this information further and show that the corrector rescales to a variant of the Gaussian free field. This is the purpose of \cite[Chapter~5]{AKMbook} (which can also serve as a good reference for the definition and basic properties of white noise and Gaussian free fields).

\smallskip

Once optimal estimates on the corrector are obtained, one can deduce optimal error estimates for general homogenization problems such as the convergence of the solution to \eqref{e.pde.eps} as $\ep$ tends to zero. Intuitively, this is made possible by the fact that the limit homogenized solution is smooth; on scales intermediate between the microscopic scale of oscillation of the coefficients and the macroscopic scale of the domain, the homogenized solution is thus well approximated by an affine function, and therefore the solution to \eqref{e.pde.eps} is well approximated by a corrector function with slope given by the local value of the gradient of the homogenized solution. Formally, this is encoded in a comparison between the solution of the heterogeneous equation and a ``two-scale expansion'' which consists in attaching the oscillations of correctors to the solution of the homogeneous problem. Optimal error estimates for such problems are proved in \cite[Chapter~6]{AKMbook}. In particular, it is shown that for sufficiently smooth domain $U$ and boundary condition $f$, and for $u_\ep$ and $\bar u$ solution to \eqref{e.pde.eps} and \eqref{e.pde.homog} respectively, we have
\begin{equation}  
\label{e.error}
\|u_\ep - \bar u\|_{L^2(U)} \ls 
\Ll\{
\begin{array}{ll}  
\ep \Ll| \log \ep \Rr|^\frac 1 2 & \quad \text{if } d = 2, \\
\ep  & \quad \text{if } d \ge 3.
\end{array}
\Rr.
\end{equation}
These estimates are sharp. (In dimension $d = 1$, it is easy to see that the difference between $u_\ep$ and $\bar u$ looks like a random walk trajectory with steps of size~$\ep$, and in particular $\|u_\ep - \bar u\|_{L^2(U)}$ is of the order of $\sqrt{\ep}$.) A precise version of the estimates in~\eqref{e.error} can be found in \cite[Theorem~6.14]{AKMbook}.

%
%
%
%
%
%
\section{Conclusion and perspectives}

Summarizing, renormalization ideas provide a powerful guiding principle towards the establishment of optimal quantitative estimates in the homogenization of problems of the form of \eqref{e.pde.eps}. The driving mechanism is a change of focus away from a direct analysis of solutions, aiming instead at a ``progressive homogenization'' of the equation itself, or more precisely, of the associated energy. 

\smallskip

Beyond the interest of the specific problem of finding rates of convergence for the homogenization of equations of the form of \eqref{e.pde.eps}, the mathematical implementation of this program is an opportunity to develop a rich set of ideas and tools that can be of interest for a variety of related questions. For example, the observation \eqref{e.small.contrast}-\eqref{e.almost.av} was influential for the development of new algorithms for the computation of the homogenized matrix $\ahom$, see \cite{efficient}. Perhaps surprisingly, the homogenized operator can also be used for the efficient computation of the solution to \eqref{e.pde.eps} at arbitrarily high precision. Indeed, the computational avatar of the idea of renormalization is the multigrid method. This method does not perform well for highly oscillating coefficients, but it was realized in \cite{AHKM,chenlin} that the homogenized operator can be used as a suitable coarse operator in such methods to restore high computational efficiency.

\smallskip

The results presented here have already been extended to a variety of other settings. For instance, elliptic equations with non-symmetric coefficients and parabolic equations with fixed or time-dependent coefficients can be treated with similar techniques, see \cite[Chapters~10 and 8]{AKMbook} and \cite{ABM}. Some significant steps have been taken towards very general non-linear equations, see \cite[Chapter~11]{AKMbook} and \cite{AFK}. The uniform ellipticity assumption \eqref{e.unif.ellip} can also be relaxed: indeed, results similar to those presented here have been obtained in the case of discrete finite-difference equations posed on a percolation cluster \cite{AD,D1}. Some results have also been obtained for problems involving differential forms \cite{D2}, as well as for a probability model of interfaces known as the Ginzburg-Landau or ``$\nabla \phi$'' model \cite{D3}. Future developments will hopefully include the analysis of certain hypoelliptic equations and interacting particle systems.

\subsection*{Acknowledgments} I would like to very warmly thank Scott Armstrong, Alexandre Bordas, Paul Dario, Chenlin Gu, Antti Hannukainen, and Tuomo Kuusi for being such outstanding collaborators. I was partially supported by the ANR grants LSD (ANR-15-CE40-0020-03) and Malin (ANR-16-CE93-0003) and by a grant from the NYU--PSL Global Alliance.

\small
\bibliographystyle{abbrv}
\bibliography{review}

\end{document}